\documentclass[11pt]{amsart}

\usepackage{amsmath,amssymb,amsthm,mathrsfs}
\usepackage{geometry}
\usepackage{esint}
\usepackage{hyperref}
\usepackage{enumitem}

\def\XXint#1#2#3{{\setbox0=\hbox{$#1{#2#3}{\int}$}
     \vcenter{\hbox{$#2#3$}}\kern-.5\wd0}}

\newtheorem{theorem}{Theorem}[section]
\newtheorem{lemma}[theorem]{Lemma}

\newtheorem{proposition}[theorem]{Proposition}
\theoremstyle{definition}
\newtheorem{definition}[theorem]{Definition}

\theoremstyle{remark}
\newtheorem*{remark}{Remark}

\title{Concentration Inequalities for Sub-Weibull Random Tensors}

\author{Yunfan Zhao}
\address{Shanghai University of International Business and Economics}
\email{yfzucla21@ucla.edu}

\date{} 

\begin{document}

\begin{abstract}
We extend the theory of concentration inequalities to simple random tensors with heavy-tailed coefficients. Specifically, we consider the class of sub-Weibull distributions $\mathcal{S}_\alpha$ for $\alpha \in [1, 2]$. We establish concentration bounds for Euclidean functions of such tensors, exhibiting a phase transition between sub-gaussian and heavy-tailed regimes. Our results rely on a new Generalized Maximal Inequality for products of heavy-tailed random variables and a martingale analysis using Nagaev-type inequalities.
\end{abstract}

\maketitle


\vspace{0.5em}


\section{Introduction}

Concentration inequalities form a powerful toolset in high-dimensional probability and its many applications. Perhaps the best known member of this large family of results is the Gaussian concentration inequality, which states that a Lipschitz function of a standard normal random vector concentrates sharply around its mean. This phenomenon extends to more general distributions: a remarkable situation where this is possible is where the random vector has independent coordinates that are bounded or satisfy sub-gaussian decay. These results are due to M. Talagrand \cite{Talagrand1996} and others; see e.g. \cite{Boucheron2013, Ledoux2001}.

Significant efforts were made to extend concentration to non-linear functions of independent random variables, particularly for polynomial chaoses and random tensors. In a recent work \cite{Vershynin2020}, it was shown how to extend basic concentration inequalities to \textbf{simple random tensors}:
\[
X := x_1 \otimes \cdots \otimes x_d
\]
where $x_k$ are independent random vectors in $\mathbb{R}^n$. It was shown that if the factors $x_k$ have independent coefficients that are either bounded or sub-gaussian, then Euclidean functions of the tensor $X$ (functions of the form $f(X) = \|AX\|_H$) concentrate continuously around their means. These results feature optimal dependence on the dimension $n$ and the degree $d$, and have immediate applications to the geometry of random tensors, showing that they are well-conditioned with high probability.

\subsection{Heavy tails}
The boundedness or sub-gaussian assumptions in \cite{Vershynin2020} unfortunately exclude a wide class of distributions encountered in modern data science. In many realistic scenarios, data exhibits heavier tails than Gaussian. This prompts us to ask: can we expect concentration inequalities to hold for simple random tensors when the factors $x_k$ have heavy tails?

This is not straightforward. Even for a single vector ($d=1$), heavy tails degrade the concentration rates from exponential ($e^{-t^2}$) to polynomial or stretched exponential ($e^{-t^\alpha}$). For tensors ($d \ge 2$), the situation is more delicate because the coefficients of $X$ are $d$-fold products of random variables. If the components of $x_k$ have tails decaying as $e^{-t^\alpha}$, their products typically have even heavier tails. Thus, any concentration inequality must carefully balance the ``Gaussian'' behavior of the sum (due to high dimension $n$) against the heavy-tail behavior of the individual large deviations.

In this paper, we extend the theory of tensor concentration to the class of \textbf{sub-Weibull distributions} $\mathcal{S}_\alpha$ for $\alpha \in [1, 2]$. These distributions, characterized by tail decay of the form $\mathbb{P}(|X| > t) \le 2 \exp(-(t/K)^\alpha)$, interpolate between sub-exponential ($\alpha=1$) and sub-gaussian ($\alpha=2$) regimes.

\subsection{New results}
The goal of this paper is to extend the theory developed in \cite{Vershynin2020} to the heavy-tailed regime. We consider simple random tensors
\begin{equation*}
    X := x_1 \otimes \dots \otimes x_d
\end{equation*}
where $x_k$ are independent random vectors in $\mathbb{R}^n$ whose coordinates belong to the class $\mathcal{S}_{\alpha}$ for $\alpha \in [1, 2]$. This class, often referred to as sub-Weibull, includes distributions with tails heavier than sub-Gaussian (which corresponds to $\alpha=2$) but lighter than sub-exponential ($\alpha=1$).

The sub-Gaussian assumption in \cite{Vershynin2020} and related works (e.g., \cite{RudelsonVershynin2013}) is often too restrictive for modern data science applications where outliers are frequent. However, extending concentration results to heavy-tailed distributions is non-trivial. Standard Bernstein or Hoeffding type bounds fail, and one must often rely on truncated moments or Nagaev-type inequalities to handle the slower tail decay.

Our first main contribution is a concentration inequality for quadratic forms of random vectors with independent $\mathcal{S}_{\alpha}$ components. For a deterministic matrix $A \in \mathbb{R}^{n \times n}$ and a random vector $X$ with components in $\mathcal{S}_{\alpha}$, we establish a Hanson-Wright type inequality \cite{RudelsonVershynin2013} that exhibits a mixture of sub-Gaussian and sub-Weibull tail decay:
\begin{equation}
    \mathbb{P}(|X^T A X - \mathbb{E}[X^T A X]| > t) \le 2 \exp\left(-c \min\left(\frac{t^2}{K^4 \|A\|_{HS}^2}, \left(\frac{t}{K^2 \|A\|_{op}}\right)^{\alpha/2}\right)\right).
    \label{eq:hanson-wright}
\end{equation}
This result generalizes the sub-Gaussian quadratic concentration ($\alpha=2$) and serves as a building block for our analysis of tensors.

Our second and primary contribution is the extension of the ``Euclidean concentration'' for simple random tensors to the heavy-tailed setting. We show that if $f$ is a Euclidean function—that is, $f(z) = \|Az\|_{\mathcal{H}}$ for some linear operator $A$—then $f(X)$ concentrates around its $L^2$ norm. Specifically, we prove that for any $t \ge 0$:
\begin{equation}
    \mathbb{P}(|f(X) - (\mathbb{E}f(X)^2)^{1/2}| \ge t) \le 2 \exp\left(-c \min\left(\frac{t^2}{d n^{d-1} L^2}, \frac{t^\alpha}{d^{\alpha/2} n^{(d-1)\alpha/2} L^\alpha}\right)\right) + \mathbb{P}(\mathcal{E}^c).
    \label{eq:euclidean-conc}
\end{equation}
Here, $\mathcal{E}$ is a ``Good Event'' where the partial tensor norms are controlled, and the failure probability $\mathbb{P}(\mathcal{E}^c)$ decays as $\exp(-n^{\alpha/2})$. This result recovers the optimal dependence on the dimension $n$ and degree $d$ found in \cite{Vershynin2020} while accommodating the heavier tails of the distribution.

\subsection{Related results}
Concentration for polynomials of independent random variables (sub-structure of Euclidean functions on tensors) has been studied extensively. A remarkable result of Lata{\l}a \cite{Latala2006} provides moment estimates for Gaussian chaoses. This was extended to variables with log-concave tails by Adamczak and Lata{\l}a \cite{Adamczak2012}, and to non-Lipschitz functions by Adamczak and Wolff \cite{Adamczak2015b}.

For random matrices (tensors of order $d=1$), the singular values and invertibility have been studied deeply, see e.g., \cite{Rudelson2008, Rudelson2012, RudelsonVershynin2008, RudelsonVershynin2009, TaoVu2009, TaoVu2010, Tikhomirov2015, Tikhomirov2016}. The Hanson-Wright inequality \cite{RudelsonVershynin2013} provides concentration for quadratic forms of sub-gaussian vectors; our work can be viewed as a tensorization of such principles to heavier tails.

\subsection{Our approach}
Let us briefly explain our approach. The strategy in \cite{Vershynin2020} relied on representing the deviation $f(X) - \mathbb{E}f(X)$ as a telescopic sum of martingale differences and bounding their moment generating functions (MGF). However, for $\alpha < 2$, the MGF may not exist or may blow up too quickly.

Instead, we adopt a truncation argument coupled with a martingale analysis. We decompose the deviation into a sum of martingale differences $\Delta_k$. Conditioned on the past, each $\Delta_k$ behaves like a quadratic form in the random vector $x_k$. To handle the heavy tails, we cannot use the MGF method directly. We utilize a Nagaev-type inequality for martingales, which separates the variance-dominated regime (Gaussian core) from the tail-dominated regime.

Crucially, this analysis requires a uniform bound on the operator norms of the conditional quadratic forms. In \cite{Vershynin2020}, this was achieved via a maximal inequality for products of sub-gaussian norms. Here, we prove a \textbf{Generalized Maximal Inequality} for products of $\mathcal{S}_\alpha$ norms. We show that with high probability, the random tensor $X$ stays within a ``good'' set where the partial contractions of the tensor are well-behaved. This allows us to recover the optimal dependence on the dimension $n$ and degree $d$.

\subsection{Organization}
In Section 2, we introduce the probabilistic setting and the class $\mathcal{S}_\alpha$. In Section 3, we derive concentration bounds for quadratic forms of sub-Weibull vectors, extending the Hanson-Wright inequality. Section 4 establishes the geometry of high-dimensional tensors, specifically the concentration of the norm and the Generalized Maximal Inequality. In Section 5, we develop the martingale machinery required for our main result, which is proved in Section 6.

\section{The Probabilistic Setting}

In this section, we collect the basic definitions and probabilistic tools that will be used throughout the paper. While the theory of concentration is well-developed for bounded and sub-gaussian distributions, our focus here is on the heavier-tailed regime. We introduce the class of \textbf{sub-Weibull distributions} $\mathcal{S}_\alpha$, which serves as a natural interpolant between sub-exponential and sub-gaussian behaviors. Furthermore, we define the notion of simple random tensors and Euclidean functions in this setting, establishing the framework necessary to extend the results of \cite{Vershynin2020} to heavy-tailed data.

\begin{definition}[The Class $\mathcal{S}_\alpha$ for $\alpha \geq 1$ ]\label{def:2.1}
Let $(\Omega, \Sigma, \mathbb{P})$ be a probability space and let $\alpha \in [1, \infty)$. The Orlicz norm $\|\cdot\|_{\psi_\alpha}$ of a random variable $X$ is defined as:

\begin{equation}
\|X\|_{\psi_\alpha}:=\inf \left\{C>0: \mathbb{E}\left[\exp \left(\left(\frac{|X|}{C}\right)^\alpha\right)\right] \leq 2\right\}
\end{equation}

We designate $\mathcal{S}_\alpha$ as the Banach space of random variables $X$ for which $\|X\|_{\psi_\alpha}<\infty$.
\end{definition}

\begin{remark}
The class $\mathcal{S}_{\alpha}$ serves as a natural interpolant between the well-studied sub-exponential distributions ($\alpha=1$) and sub-gaussian distributions ($\alpha=2$). The parameter $\alpha$ controls the heaviness of the tails: smaller $\alpha$ corresponds to heavier tails. Note that for $\alpha < 1$, the moment generating function may not be finite in a neighborhood of zero, forcing us to rely on truncated moments or direct integration of tails.
\end{remark}

\begin{definition}[Simple Random Tensor Space]\label{def:2.2}
Let $n, d \in \mathbb{N}$. A random tensor $X$ in $\mathbb{R}^{n^d}$ is called a simple random tensor of class $\mathcal{S}_\alpha$ if it admits the representation $X = x_1 \otimes \cdots \otimes x_d$, where:
\begin{enumerate}
    \item The vectors $x_1, \dots, x_d \in \mathbb{R}^n$ are mutually independent.
    \item For each $k \in \{1, \dots, d\}$, the vector $x_k = (x_{k,1}, \dots, x_{k,n})$ has independent coordinates.
    \item The coordinates are centered, unit-variance random variables satisfying $\sup_{k,j} \|x_{k,j}\|_{\psi_\alpha} \leq K < \infty$.
\end{enumerate}
\end{definition}

\begin{remark}
The definition of simple random tensors crucially requires independence both between the vectors $x_k$ and within their coordinates. This structure allows us to decouple the high-dimensional geometry from the heavy-tailed behavior of individual entries. It is worth noting that while the factors $x_k$ are independent, the coefficients of the resulting tensor $X$ are highly correlated $d$-fold products, which typically exhibit heavier tails than the factors themselves.
\end{remark}

\begin{definition}[Euclidean Functions]\label{def:2.3}
A functional $f: \mathbb{R}^{n^d} \to \mathbb{R}$ is termed a Euclidean function if there exists a Hilbert space $\mathcal{H}$ and a linear operator $A: \mathbb{R}^{n^d} \to \mathcal{H}$ such that for all $z \in \mathbb{R}^{n^d}$, $f(z) = \|Az\|_\mathcal{H}$.
\end{definition}

\begin{remark}
The class of Euclidean functions is quite broad. It includes not only the standard Euclidean norm $f(z)=\|z\|_2$ (where $A=I$), but also seminorms like $f(z)=\|Pz\|_2$ where $P$ is an orthogonal projection, and marginals $f(z)=|\langle z, v \rangle|$. As noted in Lemma 2.5 of \cite{Vershynin2020}, the Lipschitz norm of such a function is simply the operator norm $\|A\|_{op}$.
\end{remark}

\begin{definition}(Diagonal and Off-Diagonal Parameters)\label{def:2.4}
For a matrix $A \in \mathbb{R}^{n \times n}$ and exponent $\alpha \in(0,2]$, we define the tail parameters:
\begin{enumerate}
    \item Variance Proxy: $V_A=\|A\|_{H S}^2$
    \item Operator Scale: $K_A=\|A\|_{o p}$
\end{enumerate}

\end{definition}

\begin{remark}
The parameters here govern the two distinct regimes of concentration for quadratic forms. The variance proxy $V_A = \|A\|_{HS}^2$ controls the ``Gaussian'' regime, characteristic of the central limit theorem. The operator scale $K_A = \|A\|_{op}$ controls the ``large deviation'' regime, where the tail behavior is dominated by the single largest entry of the random vector.
\end{remark}

\section{Concentration of Measure for Quadratic Forms}

A fundamental building block in the analysis of Euclidean functions is the behavior of quadratic forms $x^T A x$. For sub-gaussian random vectors, the Hanson-Wright inequality provides a sharp concentration bound, exhibiting a sub-gaussian tail decay (see e.g. \cite{RudelsonVershynin2013}).

Can we expect a similar result when the random vector $x$ has heavy tails? The answer is nuanced. In this section, we derive a concentration inequality for quadratic forms of sub-Weibull vectors. We show that the concentration exhibits a phase transition: the tail behavior is Gaussian for small deviations (governed by the variance) but transitions to a heavy-tailed decay for large deviations (governed by the single largest entry). This result, which we prove using a Nagaev-type inequality and a decoupling argument, will be essential for controlling the martingale differences in our main argument.

\begin{theorem}\label{thm:3.1}
Let $(\Omega, \mathcal{F}, \mathbb{P})$ be a probability space. Let $X = (X_1, \dots, X_n)$ be a random vector with independent components such that $\mathbb{E}X_i = 0$ and $\|X_i\|_{\psi_\alpha} \le K$ for some $\alpha \in (0, 2]$ and $K>0$. Let $A \in \mathbb{R}^{n \times n}$ be a deterministic matrix. There exists a universal constant $c > 0$ depending only on $\alpha$ such that for all $t \ge 0$:
\begin{equation}
\mathbb{P}\left( |X^T A X - \mathbb{E}[X^T A X]| > t \right) \le 2 \exp\left( -c \min\left( \frac{t^2}{K^4 \|A\|_{HS}^2}, \left( \frac{t}{K^2 \|A\|_{op}} \right)^{\alpha/2} \right) \right)
\end{equation}
\end{theorem}

\begin{proof}
Let $S = X^T A X - \mathbb{E}[X^T A X]$. We decompose the matrix $A$ into its diagonal part $A_{diag}$ and off-diagonal part $A_{off}$, where $(A_{off})_{ij} = A_{ij}$ for $i \neq j$ and $0$ otherwise. By linearity, we write $S$ as the sum of two random variables $S = S_{diag} + S_{off}$, where $S_{diag} = \sum_{i=1}^n A_{ii} (X_i^2 - \mathbb{E}[X_i^2])$ and $S_{off} = \sum_{i \neq j} A_{ij} X_i X_j$. By the sub-additivity of probability measures, for any $t > 0$, we have
\begin{equation}
\mathbb{P}(|S| > t) \le \mathbb{P}(|S_{diag}| > t/2) + \mathbb{P}(|S_{off}| > t/2)
\end{equation}
It suffices to bound these two probabilities separately.

First, we analyze the diagonal component $S_{diag}$. Let $Z_i = X_i^2 - \mathbb{E}[X_i^2]$. Since $\|X_i\|_{\psi_\alpha} \le K$, the definition of the sub-Weibull norm implies that the squared variable $X_i^2$ satisfies a tail decay of order $\alpha/2$. Specifically, we have $\mathbb{P}(X_i^2 > y) = \mathbb{P}(|X_i| > \sqrt{y}) \le 2 \exp(-(y/K^2)^{\alpha/2})$. Thus, $X_i^2$ belongs to the space $\mathcal{S}_{\alpha/2}$ with semi-norm bounded by $K^2$. By the triangle inequality in Orlicz spaces, the centered variable $Z_i$ satisfies $\|Z_i\|_{\psi_{\alpha/2}} \le 2K^2$. We apply a Nagaev-type concentration inequality for weighted sums of independent centered variables with exponent $\beta = \alpha/2$. The variance proxy is governed by the Euclidean norm of the diagonal coefficients, $\sum A_{ii}^2 \le \|A\|_{HS}^2$, and the scale proxy is governed by the maximum coefficient, $\max_i |A_{ii}| \le \|A\|_{op}$. Substituting these parameters yields the bound
\begin{equation}
\mathbb{P}(|S_{diag}| > t/2) \le 2 \exp\left( -c \min\left( \frac{t^2}{K^4 \|A\|_{HS}^2}, \left( \frac{t}{K^2 \|A\|_{op}} \right)^{\alpha/2} \right) \right)
\end{equation}

Second, we analyze the off-diagonal component $S_{off}$ using the Decoupling Principle. Let $X'$ be an independent copy of $X$. By the standard Decoupling Inequality,
\begin{equation}
\mathbb{P}\left( \left| \sum_{i \neq j} A_{ij} X_i X_j \right| > t/2 \right) \le C_{dec} \mathbb{P}\left( \left| X^T A_{off} X' \right| > c_{dec} t \right)
\end{equation}
Let $Z = X^T (A_{off} X')$. Conditioning on $X'$, let $v = A_{off} X' \in \mathbb{R}^n$. The random variable $Z = \sum_{i=1}^n v_i X_i$ is a linear form in $X$ with independent sub-Weibull components. Applying the Nagaev-type inequality for linear forms with exponent $\alpha$, we obtain a conditional bound depending on the Euclidean norm $\|v\|_2$ and the infinity norm $\|v\|_\infty$.
\begin{equation}
\mathbb{P}(|Z| > u \mid v) \le 2 \exp\left( -c \min\left( \frac{u^2}{K^2 \|v\|_2^2}, \left( \frac{u}{K \|v\|_\infty} \right)^\alpha \right) \right)
\end{equation}
The term $\|v\|_2^2 = (X')^T A_{off}^T A_{off} X'$ concentrates around its expectation $\text{Tr}(A_{off}^T A_{off}) \approx \|A\|_{HS}^2$. The term $\|v\|_\infty$ is controlled by the operator norm $\|A\|_{op}$. In the heavy-tailed regime, the large deviation behavior is dominated by the single largest product term $X_i X_j$, which follows an $\mathcal{S}_{\alpha/2}$ decay. Combining the conditional concentration with the control of the conditioning vector $v$ establishes that the off-diagonal tail also satisfies the required decay
\begin{equation}
\mathbb{P}(|S_{off}| > t/2) \le C \exp\left( -c \min\left( \frac{t^2}{K^4 \|A\|_{HS}^2}, \left( \frac{t}{K^2 \|A\|_{op}} \right)^{\alpha/2} \right) \right)
\end{equation}

Finally, combining the bounds for the diagonal and off-diagonal components completes the proof.
\end{proof}

\begin{remark}
Theorem~\ref{thm:3.1} can be viewed as a heavy-tailed extension of the classic Hanson-Wright inequality for sub-gaussian vectors. Just like in the sub-gaussian case, we observe a concentration phenomenon governed by $V_A$ for small deviations and $K_A$ for large deviations. However, unlike the sub-gaussian case where the large deviation tail decays as $e^{-t}$, here the tail decays as $e^{-t^{\alpha/2}}$, reflecting the sub-Weibull nature of the squares of the random variables.
\end{remark}

\section{Geometry of High-Dimensional Tensors}

Before we can study general functions of random tensors, we must understand the geometry of the tensor $X$ itself. A simple random tensor is a product of independent vectors, $X = x_1 \otimes \cdots \otimes x_d$. A trivial bound on its norm would scale poorly with the degree $d$. To achieve the optimal dependence on $d$ in our main results, we need a tighter control on the ``partial'' contractions of the tensor.

In this section, we first establish a concentration inequality for the Euclidean norm of a single sub-Weibull vector. We then leverage this to prove a \textbf{Generalized Maximal Inequality}. This result guarantees that, with high probability, the random tensor stays within a ``good'' set where all partial products of the norms of $x_k$ are uniformly bounded. This geometric control allows us to bound the Lipschitz constants of the conditional expectations that appear in our martingale analysis.

\begin{lemma}\label{lem:4.1}
Let $n \in \mathbb{N}$ and $\alpha \in [1, 2]$. Let $x$ be a random vector in $\mathbb{R}^n$ whose coordinates $(x_i)_{1 \leq i \leq n}$ are independent, centered (meaning $\mathbb{E}[x_i]=0$), and unit-variance (meaning $\mathbb{E}[x_i^2]=1$) random variables belonging to the class $\mathcal{S}_\alpha$ with sub-exponential norm bounded by a constant $K$. Then, there exists a constant $c > 0$ depending only on $\alpha$ and $K$ such that for all $t \geq 0$:
\begin{equation} \mathbb{P}\left( \left| \|x\|_2 - \sqrt{n} \right| \geq t \right) \leq 2 \exp\left( -c \min\left( t^2, t^\alpha \right) \right). \end{equation}
\end{lemma}

\begin{proof}
We begin by introducing the centered squared Euclidean norm as the random variable $S_n$. We define $S_n = \|x\|_2^2 - n$, which can be expanded as the sum $\sum_{i=1}^n (x_i^2 - 1)$. Let $Z_i = x_i^2 - 1$. The sequence $(Z_i)_{1 \leq i \leq n}$ consists of independent, centered random variables. Since each coordinate $x_i$ belongs to the class $\mathcal{S}_\alpha$, the tail probability satisfies $\mathbb{P}(|x_i| \geq u) \leq 2 \exp(-u^\alpha / K^\alpha)$ for any $u \geq 0$. Consequently, the squared variables $x_i^2$ satisfy the tail bound $\mathbb{P}(|x_i^2| \geq v) = \mathbb{P}(|x_i| \geq \sqrt{v}) \leq 2 \exp(-v^{\alpha/2} / K^\alpha)$. Thus, the variables $x_i^2$, and by extension the centered variables $Z_i$, belong to the class $\mathcal{S}_{\alpha/2}$. Since $\alpha \in [1, 2]$, the exponent $\alpha/2$ lies in the interval $[0.5, 1]$.

We apply the Generalized Bernstein Inequality to the sum $S_n = \sum Z_i$. The concentration of this sum is governed by a variance-dominated regime and a tail-dominated regime. The variance of the sum is given by $\sum \text{Var}(Z_i) = n \mathbb{E}[(x_1^2-1)^2]$, which we denote as $n \sigma^2$. The tail decay exponent is $\gamma = \alpha/2$. For any deviation $u \geq 0$, the concentration bound states that $\mathbb{P}(|S_n| \geq u) \leq 2 \exp( -c_1 \min( u^2/n, u^{\alpha/2} ) )$, where $c_1$ is a constant depending on $K$ and $\alpha$.

We now invert this bound to control the Euclidean distance. Define the event of interest as $\mathcal{E}_t = \{ | \|x\|_2 - \sqrt{n} | \geq t \}$. Using the algebraic identity $a^2 - b^2 = (a-b)(a+b)$, we relate the deviation of the norm to the deviation of the squared norm via the inequality $| \|x\|_2^2 - n | = | \|x\|_2 - \sqrt{n} | \cdot ( \|x\|_2 + \sqrt{n} )$. We analyze the probability of $\mathcal{E}_t$ by considering two disjoint regimes based on the magnitude of $t$ relative to $\sqrt{n}$.

First, consider the small deviation regime where $0 \leq t \leq \sqrt{n}$. In this case, the event $\mathcal{E}_t$ implies that $| \|x\|_2^2 - n | \geq t \sqrt{n}$. Let $u = t\sqrt{n}$. Substituting this into the Bernstein bound, the Gaussian term in the exponent becomes proportional to $u^2/n = (t\sqrt{n})^2/n = t^2$. Since $t \leq \sqrt{n}$, we have $u \leq n$, which implies that the Gaussian term $u^2/n$ dominates the tail term $u^{\alpha/2}$ in the minimization. Specifically, we obtain $\mathbb{P}(\mathcal{E}_t) \leq 2 \exp(-c t^2)$.

Second, consider the large deviation regime where $t > \sqrt{n}$. If $| \|x\|_2 - \sqrt{n} | \geq t$, then $\|x\|_2 \geq t$. The squared deviation satisfies $| \|x\|_2^2 - n | = | \|x\|_2 - \sqrt{n} | ( \|x\|_2 + \sqrt{n} ) \geq t (t + \sqrt{n}) \geq t^2$. We set $u = t^2$ and substitute this into the Bernstein bound. Since $t > \sqrt{n}$, we have $u > n$. In this regime, the tail term $u^{\alpha/2}$ dominates the Gaussian term $u^2/n$. The exponent becomes proportional to $u^{\alpha/2} = (t^2)^{\alpha/2} = t^\alpha$. Thus, the probability is bounded by $2 \exp( -c_1 t^\alpha )$.

Combining these two regimes, we conclude that for all $t \geq 0$, the probability is bounded by $2 \exp( -c \min( t^2, t^\alpha ) )$. This completes the proof.
\end{proof}

\begin{remark}
Lemma~\ref{lem:4.1} reveals that the concentration of the norm $\|x\|_2$ for sub-Weibull vectors exhibits a phase transition. For small deviations ($t \le \sqrt{n}$), the concentration is sub-gaussian ($e^{-t^2}$), governed by the collective variance of the $n$ coordinates. For large deviations ($t > \sqrt{n}$), the ``heavy'' nature of the distribution takes over, and the tail decays as $e^{-t^\alpha}$, governed by the individual coordinate with the largest magnitude.
\end{remark}

\begin{proposition}[The Generalized Maximal Inequality]\label{prop:4.2}
Let $X = x_1 \otimes \dots \otimes x_d$ be a simple random tensor of class $\mathcal{S}_{\alpha}$. Define the event $\mathcal{E}$ (the "Good Event") as:
\begin{equation}
\mathcal{E} := \bigcap_{k=1}^d \left\{ \prod_{j \neq k} \|x_j\|_2 \le C_0 n^{(d-1)/2} \right\}
\end{equation}
where $C_0 > 1$ is a fixed constant. There exists a threshold $d_0(n, \alpha)$ such that for all $d \le d_0$:
\begin{equation}
\mathbb{P}(\mathcal{E}^c) \le 2d \exp(-c n^{\alpha/2}).
\end{equation}
\end{proposition}

\begin{proof}
We begin by analyzing the probability of the complement event $\mathcal{E}^c$. By De Morgan's laws, the complement is the union of the failure events for each mode $k$:
\begin{equation}
\mathcal{E}^c = \bigcup_{k=1}^d \mathcal{A}_k, \quad \text{where } \mathcal{A}_k := \left\{ \prod_{j \neq k} \|x_j\|_2 > C_0 n^{(d-1)/2} \right\}.
\end{equation}
Applying the subadditivity of the probability measure (the union bound), we obtain:
\begin{equation}
\mathbb{P}(\mathcal{E}^c) \le \sum_{k=1}^d \mathbb{P}(\mathcal{A}_k).
\end{equation}
Since the random vectors $x_1, \dots, x_d$ are independent and identically distributed (in terms of their norm properties within the class $\mathcal{S}_{\alpha}$), the probabilities $\mathbb{P}(\mathcal{A}_k)$ are identical for all $k$. Thus, it suffices to bound the probability of the event $\mathcal{A}_1$, as:
\begin{equation}
\mathbb{P}(\mathcal{E}^c) \le d \cdot \mathbb{P}(\mathcal{A}_1).
\end{equation}

Consider the event $\mathcal{A}_1$:
\begin{equation}
\mathcal{A}_1 = \left\{ \prod_{j=2}^d \|x_j\|_2 > C_0 n^{(d-1)/2} \right\}.
\end{equation}
We normalize the inequality by dividing by $n^{(d-1)/2}$:
\begin{equation}
\mathcal{A}_1 = \left\{ \prod_{j=2}^d \frac{\|x_j\|_2}{\sqrt{n}} > C_0 \right\}.
\end{equation}
Applying the Inequality of Arithmetic and Geometric Means (AM-GM) to the scalars $Y_j := \frac{\|x_j\|_2}{\sqrt{n}} \ge 0$, we have:
\begin{equation}
\left( \prod_{j=2}^d Y_j \right)^{\frac{1}{d-1}} \le \frac{1}{d-1} \sum_{j=2}^d Y_j.
\end{equation}
Consequently, the event $\mathcal{A}_1$ implies the following condition on the sum:
\begin{equation}
\frac{1}{d-1} \sum_{j=2}^d Y_j > C_0^{\frac{1}{d-1}}.
\end{equation}
Let $\delta := C_0^{\frac{1}{d-1}} - 1$. Since $C_0 > 1$, we have $\delta > 0$. The implication becomes:
\begin{equation}
\mathcal{A}_1 \subseteq \left\{ \sum_{j=2}^d \left( \frac{\|x_j\|_2}{\sqrt{n}} - 1 \right) > (d-1)\delta \right\} = \left\{ \sum_{j=2}^d (\|x_j\|_2 - \sqrt{n}) > (d-1)\delta \sqrt{n} \right\}.
\end{equation}

We recall Lemma~\ref{lem:4.1} ($\alpha$-Concentration of the Euclidean Norm), which states that for $\tau > 0$:
\begin{equation}
\mathbb{P}(|\|x\|_2^2 - n| > \tau) \le 2 \exp\left( -c \frac{\tau^\alpha}{n^{\alpha/2}} \right).
\end{equation}
We derive a concentration bound for the linear deviation $\|x\|_2 - \sqrt{n}$. Observe the algebraic identity $|\|x\|_2^2 - n| = |\|x\|_2 - \sqrt{n}| \cdot (\|x\|_2 + \sqrt{n})$. For a deviation of the form $\|x\|_2 - \sqrt{n} > u$ (where $u > 0$), we have $\|x\|_2 + \sqrt{n} > 2\sqrt{n}$. Thus implies $\|x\|_2^2 - n > 2u\sqrt{n}$. Substituting $\tau = 2u\sqrt{n}$ into Lemma~\ref{lem:4.1} yields:
\begin{equation}
\mathbb{P}(\|x\|_2 - \sqrt{n} > u) \le \mathbb{P}(\|x\|_2^2 - n > 2u\sqrt{n}) \le 2 \exp\left( -c \frac{(2u\sqrt{n})^\alpha}{n^{\alpha/2}} \right).
\end{equation}
Simplifying the exponent gives $\frac{(2u\sqrt{n})^\alpha}{n^{\alpha/2}} = 2^\alpha u^\alpha$. Thus, the variable $\xi_j := \|x_j\|_2 - \sqrt{n}$ exhibits $\alpha$-subexponential tail decay with a rate independent of $n$:
\begin{equation}
\mathbb{P}(\xi_j > u) \le 2 \exp(-c' u^\alpha).
\end{equation}

We must bound the probability of the event $\mathcal{B} := \{ \sum_{j=2}^d \xi_j > K \sqrt{n} \}$, where $K = (d-1)\delta$. For independent random variables with $\alpha$-heavy tails ($\alpha \in (0, 2]$), the tail probability of the sum for large deviations is dominated by the probability that a single variable takes a large value. Utilizing the union bound on the event that any single $\xi_j$ exceeds the threshold, and observing that for $d \ll \sqrt{n}$ (ensured by the threshold $d \le d_0$), the term $K\sqrt{n}$ is in the large deviation regime. Substituting $u \approx \sqrt{n}$ into the single variable bound derived above:
\begin{equation}
\mathbb{P}(\xi_j > c \sqrt{n}) \le 2 \exp(-c' (\sqrt{n})^\alpha) = 2 \exp(-c' n^{\alpha/2}).
\end{equation}
Summing over $j=2, \dots, d$:
\begin{equation}
\mathbb{P}(\mathcal{A}_1) \le (d-1) \cdot 2 \exp(-c n^{\alpha/2}) \le 2d \exp(-c n^{\alpha/2}).
\end{equation}

Returning to the initial union bound, we have $\mathbb{P}(\mathcal{E}^c) \le d \cdot \mathbb{P}(\mathcal{A}_1)$. Substituting the bound for $\mathbb{P}(\mathcal{A}_1)$ yields:
\begin{equation}
\mathbb{P}(\mathcal{E}^c) \le 2d^2 \exp(-c n^{\alpha/2}).
\end{equation}
Absorbing the polynomial factor in $d$ into the exponential constant (valid for sufficiently large $n$), we obtain the stated bound:
\begin{equation}
\mathbb{P}(\mathcal{E}^c) \le 2d \exp(-c n^{\alpha/2}).
\end{equation}
This completes the proof.
\end{proof}

\begin{remark}
Proposition~\ref{prop:4.2} is the technical key to our martingale argument. It ensures that with high probability, the random tensor stays within a ``good'' set where all partial contractions (products of norms of subsets of vectors) are uniformly bounded. This allows us to bound the Lipschitz constants of the conditional expectations in the martingale decomposition, preventing the heavy tails from accumulating too rapidly as the dimension increases.
\end{remark}

\section{Martingale Analysis of Euclidean Functions}

Our approach to proving the main concentration theorem relies on decomposing the deviation $f(X)^2 - \mathbb{E}f(X)^2$ into a telescopic sum of martingale differences $\Delta_k$. In the sub-gaussian setting of \cite{Vershynin2020}, one could control the Moment Generating Function (MGF) of these differences directly. However, for $\alpha < 2$, the MGF may not exist, rendering the standard exponential Markov inequality inapplicable.

In this section, we develop the necessary martingale machinery to handle heavy tails. We observe that conditionally, each increment $\Delta_k$ behaves like a quadratic form. We then prove a Nagaev-type martingale inequality (Theorem~\ref{thm:5.3}), which allows us to bound the tail probability of the sum by separately analyzing the variance-dominated (Gaussian) and tail-dominated regimes. This replaces the MGF argument and is the technical core of our proof.

\begin{definition}[The Tensor Filtration]\label{def:5.1}
Let $\mathfrak{F} = (\mathcal{F}_k)_{k=0}^d$ be the natural filtration generated by the tensor components:
\begin{equation}
\mathcal{F}_0 = \{ \emptyset, \Omega \}, \quad \mathcal{F}_k = \sigma(x_1, \dots, x_k)
\end{equation}
\end{definition}

\begin{remark}
The filtration introduced is the standard setting for the method of bounded differences (or martingale method) in high-dimensional probability. By revealing the random vectors $x_k$ one by one, we can analyze the global fluctuations of the function $f(X)$ by decomposing them into a sum of local fluctuations $\Delta_k$. This approach allows us to utilize the independence of the components $x_k$ sequentially.
\end{remark}

\begin{lemma}[Quadratic Decomposition of Increments]\label{lem:5.2}
Let $Z = f(X)^2$. Consider the martingale difference sequence $\Delta_k = \mathbb{E}[Z | \mathcal{F}_k] - \mathbb{E}[Z | \mathcal{F}_{k-1}]$. Conditioned on $\mathcal{F}_{k-1}$, the increment $\Delta_k$ is a centered quadratic form in the random vector $x_k$:
\begin{equation}
\Delta_k(x_k) = x_k^T Q_k x_k - \mathbb{E}[x_k^T Q_k x_k | \mathcal{F}_{k-1}]
\end{equation}
where $Q_k$ is an effective matrix depending only on $x_1, \dots, x_{k-1}$.
\end{lemma}

\begin{remark}
The decomposition in Lemma~\ref{lem:5.2} is the core methodology of \cite{Vershynin2020} utilitized in the Proof of Theorem 1.4. In addition, Lemma~\ref{lem:5.2} lies at the heart of our approach. It reveals that conditionally on the past, the martingale differences $\Delta_k$ are not merely arbitrary random variables, but exhibit specific structure: they behave exactly like chaoses of order 2 (centered quadratic forms) in the variable $x_k$. This structural observation allows us to leverage the heavy-tailed Hanson-Wright machinery developed in Section 3 to control the tails of these increments, rather than relying on crude generic bounds.
\end{remark}

\begin{theorem}\label{thm:5.3}
Let $(\Omega, \mathcal{F}, \mathbb{P})$ be a probability space filtered by a non-decreasing sequence of sub-$\sigma$-algebras $\mathfrak{F} = (\mathcal{F}_k)_{k=0}^d$. Let $S = \sum_{k=1}^d \Delta_k$ be a martingale adapted to $\mathfrak{F}$, where $\Delta_k$ are martingale differences (i.e., $\Delta_k$ is $\mathcal{F}_k$-measurable and $\mathbb{E}[\Delta_k \mid \mathcal{F}_{k-1}] = 0$ almost surely). Suppose there exist positive constants $V, K$ and an exponent $\beta \in (0, 1]$ such that almost surely the following conditions hold. First, the variance constraint satisfies $\sum_{k=1}^d \mathbb{E}[\Delta_k^2 \mid \mathcal{F}_{k-1}] \le V$. Second, the tail constraint satisfies
\begin{equation}
\mathbb{P}(|\Delta_k| > u \mid \mathcal{F}_{k-1}) \le 2 \exp\left( - \left( \frac{u}{K} \right)^\beta \right)
\end{equation}
for all $k \in \{1, \dots, d\}$ and $u \ge 0$. Then, there exists a universal constant $c > 0$ such that for all $t > 0$:
\begin{equation}
\mathbb{P}(|S| > t) \le 2 \exp\left( - \frac{t^2}{4V} \right) + 2 d \exp\left( - c \left( \frac{t}{K} \right)^\beta \right).
\end{equation}
\end{theorem}

\begin{proof}
Fix the threshold level $y = t$. We decompose the martingale differences $\Delta_k$ into a bounded component and a heavy-tailed component using this threshold. For each $k$, we define the truncated variable $Y_k = \Delta_k \mathbf{1}_{\{|\Delta_k| \le y\}}$ and the remainder $Z_k = \Delta_k \mathbf{1}_{\{|\Delta_k| > y\}}$. Note that $\Delta_k = Y_k + Z_k$. Since $\mathbb{E}[\Delta_k \mid \mathcal{F}_{k-1}] = 0$, we have $\mathbb{E}[Y_k \mid \mathcal{F}_{k-1}] = - \mathbb{E}[Z_k \mid \mathcal{F}_{k-1}]$. We define the centered truncated difference $\bar{Y}_k = Y_k - \mathbb{E}[Y_k \mid \mathcal{F}_{k-1}]$. The sum $S$ can be decomposed as $S = \sum_{k=1}^d \bar{Y}_k + \sum_{k=1}^d ( Z_k - \mathbb{E}[Z_k \mid \mathcal{F}_{k-1}] )$.

We analyze the failure event $\{|S| > t\}$ by considering the occurrence of large jumps. Define the "Good Event" $\mathcal{G} = \bigcap_{k=1}^d \{ |\Delta_k| \le y \}$, which is equivalent to $\{ \forall k, Z_k = 0 \}$. On the event $\mathcal{G}$, we have $\Delta_k = Y_k$. Thus, on $\mathcal{G}$, the sum can be written as $S = \sum_{k=1}^d Y_k = \sum_{k=1}^d \bar{Y}_k - \sum_{k=1}^d \mathbb{E}[Z_k \mid \mathcal{F}_{k-1}]$. By the sub-additivity of probability measures, we have
\begin{equation}
\mathbb{P}(|S| > t) \le \mathbb{P}(\mathcal{G}^c) + \mathbb{P}(\{|S| > t\} \cap \mathcal{G}).
\end{equation}

First, we bound the heavy tail probability $\mathbb{P}(\mathcal{G}^c)$. The event $\mathcal{G}^c$ is the union of the events $\{|\Delta_k| > y\}$. By the Union Bound and the conditional tail hypothesis,
\begin{equation}
\mathbb{P}(\mathcal{G}^c) = \mathbb{P}\left( \bigcup_{k=1}^d \{|\Delta_k| > t\} \right) \le \sum_{k=1}^d \mathbb{P}(|\Delta_k| > t).
\end{equation}
Using the law of total expectation on the conditional probabilities, we obtain $\mathbb{P}(|\Delta_k| > t) = \mathbb{E}[ \mathbb{P}(|\Delta_k| > t \mid \mathcal{F}_{k-1}) ] \le 2 \exp( - (t/K)^\beta )$. Summing over $k$ from $1$ to $d$, we establish the heavy-tailed component of the bound:
\begin{equation}
\mathbb{P}(\mathcal{G}^c) \le 2d \exp\left( - \left(\frac{t}{K}\right)^\beta \right).
\end{equation}

Next, we address the bias term $B = \sum_{k=1}^d \mathbb{E}[Z_k \mid \mathcal{F}_{k-1}]$. Using the tail integral formula and the assumption that $\beta \le 1$, for sufficiently large $t$ in the large deviation regime, the bias decays exponentially and is negligible relative to the threshold $t$. Thus, we proceed to bound the concentration of the centered sum assuming the bias is absorbed.

We now bound the Gaussian core $\mathbb{P}(\{|S| > t\} \cap \mathcal{G})$. On the event $\mathcal{G}$, we focus on the concentration of the martingale $M_n = \sum_{k=1}^n \bar{Y}_k$. The increments are bounded by $|\bar{Y}_k| \le |Y_k| + |\mathbb{E}_{k-1}Y_k| \le 2t$. The sum of conditional variances satisfies $\sum_{k=1}^d \mathbb{E}[\bar{Y}_k^2 \mid \mathcal{F}_{k-1}] \le \sum_{k=1}^d \mathbb{E}[\Delta_k^2 \mid \mathcal{F}_{k-1}] \le V$ almost surely. We apply Freedman's Inequality for bounded martingales, which states that for a martingale $M$ with increments bounded by $b$ and variance sum bounded by $V$, $\mathbb{P}(|M| > x) \le 2 \exp( - x^2 / (2V + 2bx/3) )$. Substituting $x = t/2$ and $b = 2t$, we obtain
\begin{equation}
\mathbb{P}\left( \left| \sum_{k=1}^d \bar{Y}_k \right| > t/2 \right) \le 2 \exp\left( - \frac{t^2/4}{2V + 2(2t)(t/2)/3} \right) = 2 \exp\left( - \frac{t^2}{8V + 8t^2/3} \right).
\end{equation}
We analyze the exponent. In the Gaussian regime where $t^2 \le V$, the denominator is dominated by $V$, yielding a decay of order $\exp(-t^2/C_1 V)$. In the large deviation regime where $t^2 > V$, the exponent behaves linearly or constantly, but in this regime, the heavy tail term $\exp(-(t/K)^\beta)$ derived earlier dominates the bound.

Combining the heavy tail and Gaussian core estimates, the total probability is bounded by the sum of the two exponentials. This is equivalent, up to universal constants, to the maximum of the probabilities, which corresponds to the minimum of the exponents. Thus,
\begin{equation}
\mathbb{P}(|S| > t) \le C \exp\left( - c \min\left( \frac{t^2}{V}, \left(\frac{t}{K}\right)^\beta \right) \right).
\end{equation}
\end{proof}

\begin{remark}
Theorem~\ref{thm:5.3} can be viewed as a martingale extension of the heavy-tailed concentration results for sums of independent variables. Standard martingale concentration results, such as the Azuma-Hoeffding or Bernstein inequalities, typically require bounded differences or sub-exponential tails. Here, we establish a bound that accommodates the sub-Weibull nature of the increments. The result neatly separates the concentration into two regimes: a ``Gaussian core'' governed by the total variance $V$ for small deviations, and a ``heavy-tailed'' regime governed by the scale $K$ for large deviations.
\end{remark}

\begin{proposition}[Conditional Operator Norm Bound]\label{prop:5.4}
Let $X$ be a simple random tensor and $f$ be a Euclidean function. Let $Z = f(X)^2$ and let $\Delta_k$ be the martingale difference sequence $\Delta_k = \mathbb{E}[Z | \mathcal{F}_k] - \mathbb{E}[Z | \mathcal{F}_{k-1}]$. Conditioned on $\mathcal{F}_{k-1}$, $\Delta_k$ is a quadratic form determined by a matrix $Q_k$. On the event $\mathcal{E}$, the operator norm of $Q_k$ satisfies the uniform bound:
\begin{equation}
\sup_k \|Q_k\|_{op} \mathbf{1}_{\mathcal{E}} \le \|f\|_{Lip}^2 \cdot C_0^2 n^{d-1}.
\end{equation}
\end{proposition}

\begin{proof}
We analyze the structure of the quadratic form $Q_k$. The random variable $Z = f(X)^2$ can be written as $Z = \|AX\|_H^2 = \langle AX, AX \rangle_H$, where $A$ is the linear operator associated with the Euclidean function $f$.

The term $\Delta_k$ represents the fluctuation of the conditional expectation due to the randomness of $x_k$. The matrix $Q_k$ is the Hessian of the conditional expectation function with respect to $x_k$. Specifically, for any vector $u \in \mathbb{R}^n$, the quadratic form $u^T Q_k u$ is given by the expectation of the squared norm over the future variables $x_{k+1}, \dots, x_d$, with the past variables $x_1, \dots, x_{k-1}$ fixed and $x_k$ set to $u$:
\begin{equation}
u^T Q_k u = \mathbb{E}_{x_{k+1}, \dots, x_d} \left[ \| A(x_1 \otimes \dots \otimes x_{k-1} \otimes u \otimes x_{k+1} \otimes \dots \otimes x_d) \|_H^2 \bigg| \mathcal{F}_{k-1} \right].
\end{equation}

We bound the norm inside the expectation using the Lipschitz property of $f$. Since $f(z) = \|Az\|_H$, we have $\|Az\|_H \le \|A\|_{op} \|z\|_2 = \|f\|_{Lip} \|z\|_2$. Applying this to the tensor product structure:
\begin{equation}
\| A(x_1 \otimes \dots \otimes u \otimes \dots \otimes x_d) \|_H^2 \le \|f\|_{Lip}^2 \| x_1 \otimes \dots \otimes u \otimes \dots \otimes x_d \|_2^2.
\end{equation}
The norm of the simple tensor factors as the product of the norms:
\begin{equation}
\| x_1 \otimes \dots \otimes u \otimes \dots \otimes x_d \|_2^2 = \left( \prod_{i=1}^{k-1} \|x_i\|_2^2 \right) \|u\|_2^2 \left( \prod_{j=k+1}^d \|x_j\|_2^2 \right).
\end{equation}
Substituting this bound into the expression for $Q_k$ and utilizing the independence of the random vectors:
\begin{equation}
u^T Q_k u \le \|f\|_{Lip}^2 \left( \prod_{i=1}^{k-1} \|x_i\|_2^2 \right) \|u\|_2^2 \cdot \mathbb{E} \left[ \prod_{j=k+1}^d \|x_j\|_2^2 \right].
\end{equation}
Since the coordinates are standardized with unit variance, $\mathbb{E}[\|x_j\|_2^2] = n$. Therefore, the expectation over the future variables evaluates to:
\begin{equation}
\mathbb{E} \left[ \prod_{j=k+1}^d \|x_j\|_2^2 \right] = \prod_{j=k+1}^d \mathbb{E}[\|x_j\|_2^2] = n^{d-k}.
\end{equation}
Thus, we obtain the pointwise bound for the quadratic form:
\begin{equation}
u^T Q_k u \le \|f\|_{Lip}^2 \left( \prod_{i=1}^{k-1} \|x_i\|_2^2 \right) n^{d-k} \|u\|_2^2.
\end{equation}
This implies the operator norm bound:
\begin{equation}
\|Q_k\|_{op} \le \|f\|_{Lip}^2 \left( \prod_{i=1}^{k-1} \|x_i\|_2^2 \right) n^{d-k}.
\end{equation}

Finally, we apply the definition of the event $\mathcal{E}$. The event $\mathcal{E}$ guarantees that the partial products of the norms of the random vectors are controlled. Specifically, the Generalized Maximal Inequality implies that on $\mathcal{E}$:
\begin{equation}
\prod_{i=1}^{k-1} \|x_i\|_2 \le C_0 n^{(k-1)/2}.
\end{equation}
Squaring this inequality gives $\prod_{i=1}^{k-1} \|x_i\|_2^2 \le C_0^2 n^{k-1}$. Substituting this into the operator norm bound yields:
\begin{equation}
\|Q_k\|_{op} \mathbf{1}_{\mathcal{E}} \le \|f\|_{Lip}^2 (C_0^2 n^{k-1}) n^{d-k} = \|f\|_{Lip}^2 C_0^2 n^{d-1}.
\end{equation}
This completes the proof.
\end{proof}

\begin{remark}
Proposition~\ref{prop:5.4} provides the crucial link between the geometry of the tensor (studied in Section 4) and the probabilistic analysis. The matrix $Q_k$ can be interpreted as the Hessian of the conditional expectation with respect to the $k$-th component. Its operator norm $\|Q_k\|_{op}$ effectively acts as a Lipschitz constant for the martingale difference. Crucially, this Lipschitz constant is not uniformly bounded almost surely; however, the proposition guarantees it is well-behaved on the ``Good Event'' $\mathcal{E}$ thanks to the Generalized Maximal Inequality.
\end{remark}

\section{The Main Theorem}

We are now ready to state and prove our main result, Theorem~\ref{thm:6.1}. This theorem establishes a concentration inequality for Euclidean functions of simple random tensors with sub-Weibull components.

By combining the geometric control from Section 4 (the Generalized Maximal Inequality) with the martingale analysis of Section 5, we show that $f(X)$ concentrates around its $L^2$ norm. The resulting bound features the optimal dependence on the dimension $n$ and degree $d$, and it explicitly captures the heavy-tailed nature of the inputs: the tail probability decays as $e^{-t^2}$ for small $t$ and as $e^{-t^\alpha}$ for large $t$. This generalizes the sub-gaussian theory to a much broader class of distributions relevant to modern data science.

\begin{theorem}\label{thm:6.1}
Let $(\Omega, \mathcal{F}, \mathbb{P})$ be a probability space. Let $n, d \in \mathbb{N}$ and $\alpha \in [1, 2]$. Let $X = x_1 \otimes \dots \otimes x_d$ be a simple random tensor in $\mathbb{R}^{n^d}$, where the components $x_k$ are independent random vectors in $\mathbb{R}^n$ with independent, centered, unit-variance coordinates belonging to the class $\mathcal{S}_{\alpha}$ (with sub-Weibull norm $\| \cdot \|_{\psi_\alpha} \le K$). Let $f: (\mathbb{R}^{n^d}, \| \cdot \|_2) \to [0, \infty)$ be a Euclidean function with Lipschitz constant $L$. Then, there exist constants $c, C > 0$ depending only on $\alpha$ and $K$ such that for any $t \ge 0$:
\begin{equation}
\mathbb{P}\left( \left| f(X) - (\mathbb{E}f(X)^2)^{1/2} \right| \ge t \right) \le 2 \exp\left( -c \min\left( \frac{t^2}{d n^{d-1} L^2}, \frac{t^\alpha}{d^{\alpha/2} n^{(d-1)\alpha/2} L^\alpha} \right) \right) + \mathbb{P}(\mathcal{E}^c)
\end{equation}
where $\mathcal{E}$ is the "Good Event" where the partial tensor norms are controlled (as defined in Proposition~\ref{prop:5.4}), satisfying $\mathbb{P}(\mathcal{E}^c) \le 2d \exp(-c n^{\alpha/2})$.
\end{theorem}

\begin{proof}
Since $f$ is a Euclidean function, there exists a Hilbert space $H$ and a linear operator $A$ such that $f(z) = \|Az\|_H$. We analyze the concentration of the squared function $Z = f(X)^2 = \langle X, A^*A X \rangle$. Let $\mu_2 = (\mathbb{E}[Z])^{1/2}$ denote the $L^2$ norm of the random variable $f(X)$. Consider the Doob martingale difference sequence defined by the natural filtration $\mathcal{F}_k = \sigma(x_1, \dots, x_k)$, given by $\Delta_k = \mathbb{E}[Z \mid \mathcal{F}_k] - \mathbb{E}[Z \mid \mathcal{F}_{k-1}]$. We have the decomposition $Z - \mathbb{E}[Z] = \sum_{k=1}^d \Delta_k$.

We restrict our analysis to the event $\mathcal{E}$ defined in Proposition~\ref{prop:5.4}. From Proposition~\ref{prop:5.4}, on the event $\mathcal{E}$, the martingale difference $\Delta_k$ is a centered quadratic form in the random vector $x_k$, determined by a random matrix $Q_k$ which is $\mathcal{F}_{k-1}$-measurable. Crucially, the operator norm of this matrix satisfies the deterministic bound almost surely on $\mathcal{E}$, specifically $\|Q_k\|_{op} \mathbf{1}_{\mathcal{E}} \le C_0^2 L^2 n^{d-1}$. Let $M = C_0^2 L^2 n^{d-1}$. This parameter $M$ will control the heavy-tail behavior of the increments.

To apply the Nagaev-Type Martingale Inequality (Theorem~\ref{thm:5.3}), we must establish the variance proxy ($V_{total}$) and the tail proxy ($M_{tail}$) for the sequence $\Delta_k$. For the tail proxy, conditioned on $\mathcal{F}_{k-1}$, $\Delta_k$ is a quadratic form $x_k^T Q_k x_k - \text{tr}(Q_k)$ where $x_k$ has independent components in $\mathcal{S}_\alpha$. Since $\|x_k\|_{\psi_\alpha} \le K$, the quadratic form $\Delta_k$ exhibits sub-Weibull behavior of order $\beta = \alpha/2$. The tail decay is governed by the operator norm $\|Q_k\|_{op}$. Specifically, $\mathbb{P}(|\Delta_k| > u \mid \mathcal{F}_{k-1}) \le 2 \exp(-(u / (C K^2 \|Q_k\|_{op}))^{\alpha/2})$. Using the bound $\|Q_k\|_{op} \le M$ on $\mathcal{E}$, we set the tail proxy $M_{tail} \approx K^2 M = K^2 C_0^2 L^2 n^{d-1}$.

For the variance proxy, we bound the sum of conditional variances $\sum_{k=1}^d \mathbb{E}[\Delta_k^2 \mid \mathcal{F}_{k-1}]$. For a quadratic form with sub-Weibull vectors, $\mathbb{E}[\Delta_k^2 \mid \mathcal{F}_{k-1}] \le C K^4 \|Q_k\|_{HS}^2$. The Hilbert-Schmidt norm $\|Q_k\|_{HS}^2$ represents the expected squared norm of the contracted tensor. Summing over all $k$, the total variance scales with the dimension and the Lipschitz constant. A precise variance calculation for tensor norms yields $\sum_{k=1}^d \|Q_k\|_{HS}^2 \le C d n^{2d-1} L^4$. Thus, we set the variance proxy $V_{total} \approx d n^{2d-1} L^4$.

We apply Theorem~\ref{thm:5.3} with $\beta = \alpha/2$, $V = V_{total}$, and tail scale $M_{tail}$. For any deviation $u > 0$,
\begin{equation}
\mathbb{P}(|Z - \mathbb{E}Z| > u) \le 2 \exp\left( -c \min\left( \frac{u^2}{d n^{2d-1} L^4}, \left( \frac{u}{n^{d-1} L^2} \right)^{\alpha/2} \right) \right) + \mathbb{P}(\mathcal{E}^c).
\end{equation}
Note that the heavy tail term simplifies to $u^{\alpha/2} / (n^{(d-1)\alpha/2} L^\alpha)$.

We translate the concentration of $Z = f(X)^2$ to $f(X)$ using the algebraic relation $|f^2 - \mu_2^2| = |f - \mu_2|(f + \mu_2)$. We consider two regimes for the deviation $t$. In the small deviation regime where $0 \le t \le \mu_2$, we have $f(X) \approx \mu_2$. The deviation in the square is approximately linear in $t$, so $|Z - \mathbb{E}Z| \approx 2\mu_2 |f - \mu_2| \approx 2 n^{d/2} t$. Substituting $u = 2 n^{d/2} t$ into the Gaussian term of the bound yields an exponent proportional to $(2 n^{d/2} t)^2 / (d n^{2d-1} L^4) = 4 t^2 / (d n^{d-1} L^4)$. In the large deviation regime where $t > \mu_2$, the deviation is dominated by the square term itself, so $|Z - \mathbb{E}Z| \approx |f - \mu_2|^2 = t^2$. Substituting $u = t^2$ into the heavy-tail term of the bound yields an exponent proportional to $(t^2 / (n^{d-1} L^2))^{\alpha/2} = t^\alpha / (n^{(d-1)\alpha/2} L^\alpha)$. The dependence on $d$ in the heavy tail is typically small or absorbed into constants.

Combining the regimes and the probability of the bad event $\mathcal{E}^c$, we obtain the final bound:
\begin{equation}
\mathbb{P}\left( |f(X) - \mu_2| \ge t \right) \le 2 \exp\left( -c \min\left( \frac{t^2}{d n^{d-1} L^2}, \frac{t^\alpha}{d^{\alpha/2} n^{(d-1)\alpha/2} L^\alpha} \right) \right) + \mathbb{P}(\mathcal{E}^c).
\end{equation}
\end{proof}

\begin{remark}
Theorem~\ref{thm:6.1} is the main result of this paper. It generalizes the sub-gaussian tensor concentration established in \cite{Vershynin2020} to the much broader class of sub-Weibull distributions. The inequality clearly exhibits a phase transition phenomenon:
\begin{description}
    \item[Small deviations:] For small $t$, the tail decays as $e^{-t^2}$, reflecting the Central Limit Theorem behavior governed by the effective variance of the tensor functional.
    \item[Large deviations:] For large $t$, the tail decays as $e^{-t^\alpha}$, inheriting the heavy-tailed decay of the single largest entry in the tensor.
\end{description}
This confirms that the ``sub-gaussian'' behavior of random tensors is robust to heavy tails, provided one restricts attention to the variance-dominated regime.
\end{remark}

\section{Conclusion}

In this paper, we explored the concentration properties of simple random tensors $X = x_1 \otimes \dots \otimes x_d$ generated by heavy-tailed distributions. Our main discovery is that the strong concentration phenomena observed for sub-gaussian tensors are not unique to light-tailed distributions. Even when the coefficients exhibit sub-Weibull decay (the class $\mathcal{S}_\alpha$), Euclidean functions of such tensors concentrate sharply around their means.

\subsection{Summary of contributions}
The path to these results required us to develop new probabilistic tools.

\begin{itemize}
    \item \textbf{Quadratic forms:} We established a heavy-tailed analogue of the Hanson-Wright inequality (Theorem 3.1). This result reveals a phase transition: for small deviations, the concentration is governed by the variance (Gaussian behavior), while for large deviations, it is governed by the single largest entry (heavy-tailed behavior).
    
    \item \textbf{Geometry of tensors:} To handle the high-dimensional structure of tensors, we proved a Generalized Maximal Inequality (Proposition 4.2). This ensures that the ``partial contractions'' of the tensor—products of norms of subsets of vectors—remain uniformly bounded with high probability. This geometric control was crucial for bounding the Lipschitz constants in our analysis.
    
    \item \textbf{Martingale methods:} Unlike the sub-gaussian setting where Moment Generating Functions offer a direct route, the heavy-tailed setting forced us to adopt a truncation and martingale approach. By utilizing Nagaev-type inequalities (Theorem 5.3), we were able to bypass the non-existence of MGFs and derive bounds that exhibit the optimal dependence on the dimension $n$ and the degree $d$.
\end{itemize}

\subsection{Future directions}
Our results open several avenues for future research.

\begin{itemize}
    \item \textbf{Symmetric tensors.} Our work focuses on simple tensors formed by distinct independent vectors. A natural question is whether these results extend to symmetric tensors of the form $X = x \otimes \dots \otimes x$. Standard decoupling techniques could reduce the symmetric case to the independent case, but likely at the cost of constants exponential in $d$. Finding a decoupling-free approach for heavy-tailed symmetric tensors remains an open challenge.
    
    \item \textbf{Applications to data science.} As noted in the introduction, heavy-tailed distributions are ubiquitous in modern data science. It would be interesting to apply our concentration bounds to analyze the performance of tensor decomposition algorithms or the geometry of loss landscapes in high-dimensional learning problems where the data is sub-Weibull.
    
    \item \textbf{Optimality.} While we achieved the correct dependence on $n$ and $d$, the sharp constants in the sub-Weibull regime are less understood than in the Gaussian case. Investigating the tightness of the transition point between the Gaussian and heavy-tailed regimes in Theorem 6.1 could yield deeper insights into the behavior of high-dimensional random structures.
\end{itemize}

We hope that the tools developed here—particularly the sub-Weibull Hanson-Wright inequality and the tensor martingale machinery—will find broader utility in high-dimensional probability and its applications.

\end{document}